\documentclass[prl,twocolumn,showpacs]{revtex4}

\usepackage{graphicx}
\usepackage{dcolumn}

\usepackage{eucal}
\usepackage[dvips]{epsfig}
\usepackage{amssymb}

\usepackage{amsmath}

\newcommand{\be}{\begin{eqnarray}}
\newcommand{\ee}{\end{eqnarray}}
\newcommand{\bse}{\begin{subequations}}
\newcommand{\ese}{\end{subequations}}

\newcommand{\bnum}{\begin{enumerate}}
\newcommand{\enum}{\end{enumerate}}

\newcommand{\bit}{\begin{itemize}}
\newcommand{\eit}{\end{itemize}}

\newcommand{\bc}{\begin{cases}}
\newcommand{\ec}{\end{cases}}

\newcommand{\bpm}{\begin{pmatrix}}
\newcommand{\epm}{\end{pmatrix}}

\newcommand{\bvm}{\begin{vmatrix}}
\newcommand{\evm}{\end{vmatrix}}

\newcommand{\mbf}{\mathbf}
\newcommand{\mbb}{\mathbb}
\newcommand{\mcal}{\mathcal}

\newcommand{\mrm}{\mathrm}

\newcommand{\vol}{\mrm{vol}}

\newcommand{\gb}{\beta}

\newcommand{\gd}{\delta}

\newcommand{\gl}{\lambda}

\newcommand{\go}{\omega}

\newcommand{\Go}{\Omega}
\newcommand{\Gc}{\Gamma}

\newcommand{\p}{\partial}
\newcommand{\f}{\frac}
\newcommand{\diff}{\mrm{d}}

\newcommand{\R}{\mathbb{R}}

\newcommand{\Z}{\mathbb{Z}}

\newcommand{\lan}{\langle}
\newcommand{\ran}{\rangle}

\newcommand{\kB}{k_B}
\newcommand{\kBT}{k_B T}

\begin{document}

\title{Non-analytic microscopic phase transitions and temperature oscillations in the microcanonical ensemble: An exactly solvable $1d$-model for evaporation}

\author{Stefan Hilbert}
 \email{hilbert@mpa-garching.mpg.de}
\affiliation{Institute for Physics, 
Humboldt-Universit\"at zu Berlin, 
Newton-Stra\ss e 15,
D-12489 Berlin,
Germany}
\author{J\"orn Dunkel} 
\email{joern.dunkel@physik.uni-augsburg.de}
\affiliation{Institute for Physics, Universit\"at Augsburg, Universit\"atsstra{\ss}e 1, D-86135 Augsburg,
Germany}

\date{\today}

\begin{abstract}
We calculate exactly both the microcanonical and canonical
thermodynamic functions (TDFs) for a one-dimensional model system with
piecewise constant Lennard-Jones type pair interactions. In the case
of an isolated $N$-particle system, the microcanonical TDFs exhibit
$(N-1)$ singular (non-analytic) microscopic phase transitions of the
formal order $N/2$,
separating $N$ energetically different evaporation (dissociation) states.
In a suitably designed evaporation experiment, these types of phase
transitions should manifest themselves in the form of pressure and temperature
oscillations, indicating cooling by evaporation. 
In the presence of a heat bath (thermostat), such oscillations are
absent, but the canonical heat capacity shows a characteristic peak, indicating
the temperature-induced dissociation of the one-dimensional chain. The
distribution of complex zeros (DOZ) of the canonical partition may be used
to identify different degrees of dissociation in the canonical ensemble.
\end{abstract}


\maketitle


\section{Introduction}
Many macroscopic systems exhibit sudden variations of
their physical properties (elasticity, conductivity, etc.), when one or more
external control parameters (e.g., energy $E$, temperature $T$) pass certain
critical values \cite{Ka67,LaLi5}. Such phenomena are usually referred to as
phase transitions (PTs). Seminal contributions to the theory of PTs are due to, e.g., Mayer et al.~\cite{Ma37,MaAc37,StMa39}, Onsager~\cite{on44},
van Hove~\cite{Ho50},
Yang and Lee  \cite{YaLe52,LeYa52}, Fisher \cite{Fi71}, Grossmann et
al. \cite{GrRo67,GrRo69,GrLe69},  Burkhardt \cite{Bu81}, Pettini et
al. \cite{CaCoPe02,FrPe04,AnEtAl05} and Cuesta and S\'anchez
\cite{CuSa04}. These authors have studied in detail canonical ensembles (CEs) and grandcanonical ensembles, corresponding to systems in contact with heat bath and particle reservoirs.
\par 
Singular, or equivalently \emph{non-analytic}, PTs are indicated by a discontinuity in the thermodynamic functions (TDFs) or one of their derivatives~\cite{Ru99}. Sometimes it is also useful to consider \emph{smooth} PTs, characterized by a strong but analytic variation in the TDFs~\cite{BoMuHa00,MuStBo01,JaKe01,AlFeHa02}.
In the presence of a heat bath, singular canonical PTs can occur in the thermodynamic limit
only~\cite{YaLe52,LeYa52,GrRo67,GrRo69,GrLe69,Fi71}, whereas finite canonical systems may, at best, exhibit smooth PTs~\cite{BoMuHa00,MuStBo01,JaKe01,AlFeHa02}.
However, the situation changes if the system under consideration is thermally isolated, corresponding to a microcanonical ensemble (MCE). Due to the different physical conditions underlying MCE and CE, respectively, one can obtain significantly different predictions for several observable quantities~\cite{WaBe94,WaDo95,Gr97,ChDuGu00,Gr01,CoEtAl05}. For example, in certain cases, microcanonical heat 
capacities can also be negative (e.g. in self-gravitating systems) whereas canonical heat capacities are generally positive. In particular, as will also be shown below,  the microcanonical TDFs of \emph{finite} isolated systems may exhibit non-analyticities. These singularities reflect evaporation/dissociation phenomena and may be interpreted as \emph{microscopic} PTs in the small system~\cite{DuHi05}.
\par 
The main objective of this article is to exemplify the differences between the
MCE and CE and to elucidate particular observational consequences by
means of a $1d$-model for evaporation. Remarkably, this model system analyzed below allows
for calculating exactly both the canonical and microcanonical TDFs for an
arbitrary number of particles. The knowledge of the exact TDFs for {\em both}
ensembles provides the basis for a detailed comparison of observables. Our main
results can be summarized as follows:
\par 
In the case of the MCE the model system
exhibits \mbox{$(N-1)$} singular \emph{microscopic} PTs, reflected by non-analytic kinks in the caloric curve $T(E)$ and the pressure curve $P(E)$ at certain critical
energy values $E_k$. The values $E_k$ can be identified with the binding energy
of different dissociation states; i.e., the singularities
(non-analyticities) separate energetically
different evaporation phases. These microscopic PTs are accompanied by strong temperature oscillations; i.e., the temperature of the system decreases when increasing the energy in the vicinity of the critical values $E_k$. This effect corresponds to cooling by evaporation (or dissociation). By contrast, non-analytic transitions are absent in the corresponding CE; i.e., if the system is embedded into
a heat bath of temperature $T$. Nevertheless, a smooth PT is observed
that also persists in the thermodynamic limit -- even though the existence of a
singular macroscopic PT is excluded by the (generalized) van Hove
theorem~\cite{Ho50,CuSa04}. 
Finally, our study of the distribution  of complex zeros (DOZ) of the canonical
partition function \cite{BoMuHa00,MuStBo01,JaKe01,AlFeHa02} suggests that the
DOZ may be used to identify different degrees of dissociation in the CE. 

\section{The model}
We consider a one-dimensional model system corresponding to
$N$ identical point particles confined by a one-dimensional box of size
$L$. The Hamiltonian reads
\begin{equation}
\label{e:Hamiltonian}
H(\mbf{p},\mbf{q};L,N)=\f{\mbf{p}^2}{2m} + U(\mbf{q};L,N)=E,
\end{equation}
with \mbox{$\mbf{q}=(q_1,\ldots, q_N)$} denoting the coordinates and 
\mbox{$\mbf{p}=(p_1,\ldots, p_N)$} the conjugate momenta. In the case of an
isolated system the total energy $E$ is conserved. The potential energy
\mbox{$U=U_\mrm{int} + U_\mrm{box}$} is determined by the interaction potential
\bse
\begin{equation}
\label{e:Uint}
U_\mrm{int}(\mbf{q};N)=\f{1}{2}\sum_{{i,j=1}\atop{i\neq j}}^N
U_\mrm{pair}(|q_i-q_j|) 
\end{equation}
and the box potential
\begin{equation}
\label{e:Ubox}
U_\mrm{box}(\mbf{q};L,N)=
\begin{cases}
0,& \mbf{q}\in [0,L]^N,\\
+\infty,&  \mrm{otherwise.}
\end{cases}
\end{equation}
The pair potential is given by
\begin{equation}
\label{e:Upair}
U_\mrm{pair}(r) = \begin{cases}
\infty, & r\leq d_\mrm{hc},\\
-U_0,& d_\mrm{hc}<r<d_\mrm{hc}+r_0,\\
0, & r\geq d_\mrm{hc}+r_0,
\end{cases}
\end{equation}
\ese
where \mbox{$d_\mrm{hc}>0$} is the hard-core diameter of a particle with respect
to pair interactions. The interaction potential \eqref{e:Upair} can be viewed as a simplified Lennard-Jones potential. The depth of the potential well is determined by the binding energy parameter \mbox{$U_0>0$} and the interaction range by the parameter $r_0$, where we shall additionally impose that 
$$
0<r_0\leq d_\mrm{hc}.
$$ 
The latter condition ensures that particles may
interact with their nearest neighbors only. Furthermore, we assume that
\mbox{$L>L_\mrm{min}\equiv(N-1)(d_\mrm{hc}+r_0)$}, i.e., the volume is assumed to be sufficiently large for realizing the completely dissociated state, corresponding to $U=0$. 
The energy $E$ of the system can take values between the ground state energy $$E_0\equiv-(N-1)U_0$$ 
and infinity. 

\section{Microcanonical ensemble}
The microcanonical ensemble (MCE) refers to an isolated system.
Thence, the control parameters are energy $E$, volume $L$ and particle number
$N$. The thermodynamic (Hertz) entropy of the MCE is given by
\cite{He10,He10a,Be67,DuHi05} 
\bse
\be
S(E,L,N)=\kB\ln \Omega(E,L,N),
\ee
where $\kB$ is the Boltzmann constant, and 
\be
\Omega(E,L,N)
=\label{e:omega}
\f{1}{N!\,h^N}
\int_{\R^N}\diff \mbf q\int_{\R^N} \diff \mbf p \;
\Theta(E-H)
\ee
\ese
the phase volume ($h$ is Planck's constant, and  $\Theta(x)\equiv 0$ for
$x<0$ and $\Theta(x)\equiv 1$ for $x\ge0$). 
Using $N$-dimensional spherical momentum  coordinates, one can rewrite
Eq. \eqref{e:omega} as  
\bse
\be
\Omega
&=&\label{e:omega-1}
C(N)\int_{\R^N}\diff \mbf q\;
(E-U)^{N/2}\;\Theta(E-U),\quad\\
C(N)&\equiv&
\f{2\,(2\pi m)^{N/2}}{\Gamma(N/2)\,N!\,N\,h^N}
\ee
\ese
where $\Gc$ denotes the Euler gamma function. 
For Hamiltonian \eqref{e:Hamiltonian} one can
calculate integral \eqref{e:omega-1} exactly, yielding (see Appendix)
\bse\label{e:result_omega}
\begin{equation}
\label{e:result_omega-a}
\Omega=
C\sum\limits_{k=0}^{N-1}
\omega_k\,(E+k U_0)^{N/2}\,\Theta(E+k U_0),
\end{equation}
where, for \mbox{$L>(N-1)(r_0+d_\mrm{hc})$},
\begin{multline}
\label{e:result_omega-b}
\omega_k(N,L)=
\binom{N-1}{k}\sum_{i=0}^{k} \binom{k}{i}\,{\left(-1 \right)}^i\,\times\\
\times\,{\left[L-(N-1)d_\mrm{hc}-r_0\left(N-1-k+i\right)\right]}^N\;.
\end{multline}
\ese
Given Eqs. \eqref{e:result_omega}, the microcanonical temperature $T$ and
pressure $P$ are obtained from the standard definitions
\mbox{$T^{-1}\equiv{\p S}/{\p E}$ and ${P}/{T}\equiv{\p S}/{\p L}$}
\cite{Hu63,Be67,Mu69,LaLi5}. For example, for the temperature one finds
\be\label{e:caloric}
\kBT=\f{2}{N}
\f{\sum\limits_{k=0}^{N-1}\omega_k\;(E+k U_0)^{N/2}\,\Theta(E+k U_0)}
{\sum\limits_{k=0}^{N-1}\omega_k\;(E+k U_0)^{N/2-1}\,\Theta(E+k U_0)},
\ee 
which reduces to the ideal gas law $E=N\kBT/2$ in the limit $E\gg N U_0$. It is
worthwhile to recall that, for a Hamiltonian of the form
\eqref{e:Hamiltonian}, the thermal energy \eqref{e:caloric} derived from the Hertz entropy is directly related to the microcanonical mean kinetic energy per degree of freedom by virtue of the equipartition theorem~\cite{Hu63,Be67}:
\be\label{e:average}
\f{\kB T}{2}=\biggl\lan \f{p_i^2}{2m}\biggr\ran,\qquad i=1,\ldots,N,
\ee 
where $\lan\,\cdot\,\ran$ denotes the average with respect to the
microcanonical probability density function 
\be\label{e:pdf}
f(q,p)=\left(\f{\p\Go}{\p E}\right)^{-1}\f{1}{N!\,h^N}\;\gd(E-H(q,p)).
\ee
Hence, for isolated ergodic systems with an arbitrary particle number $N$, the
caloric law $T(E)$ can be measured experimentally by monitoring the
kinetic energy over a sufficiently long time interval (at fixed energy
values $E$).
\par
\begin{figure}[tb]
\centerline{\includegraphics[width=1\linewidth]
{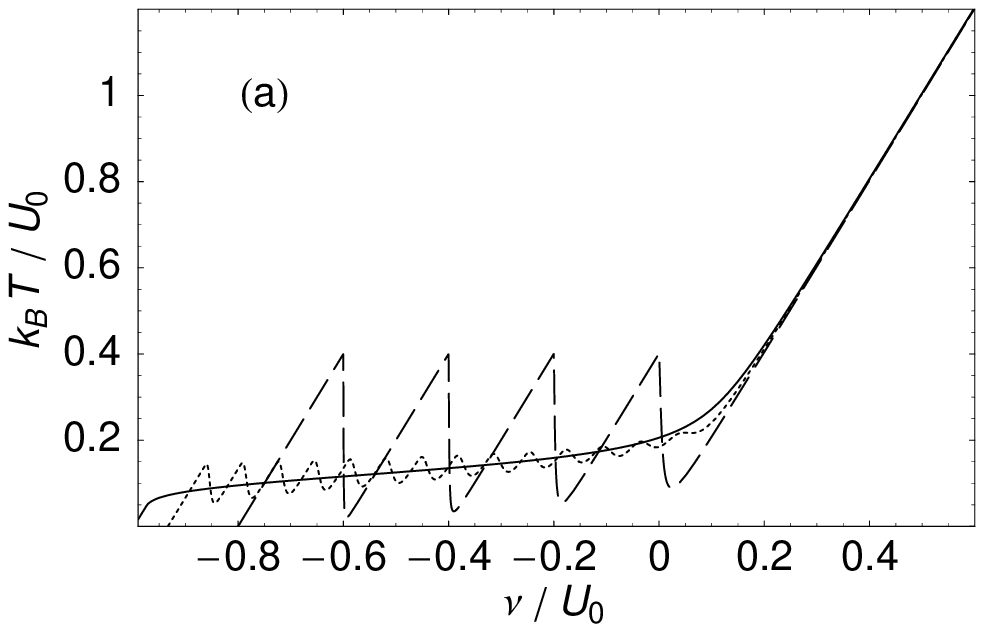}}
\centerline{\includegraphics[width=1\linewidth]
{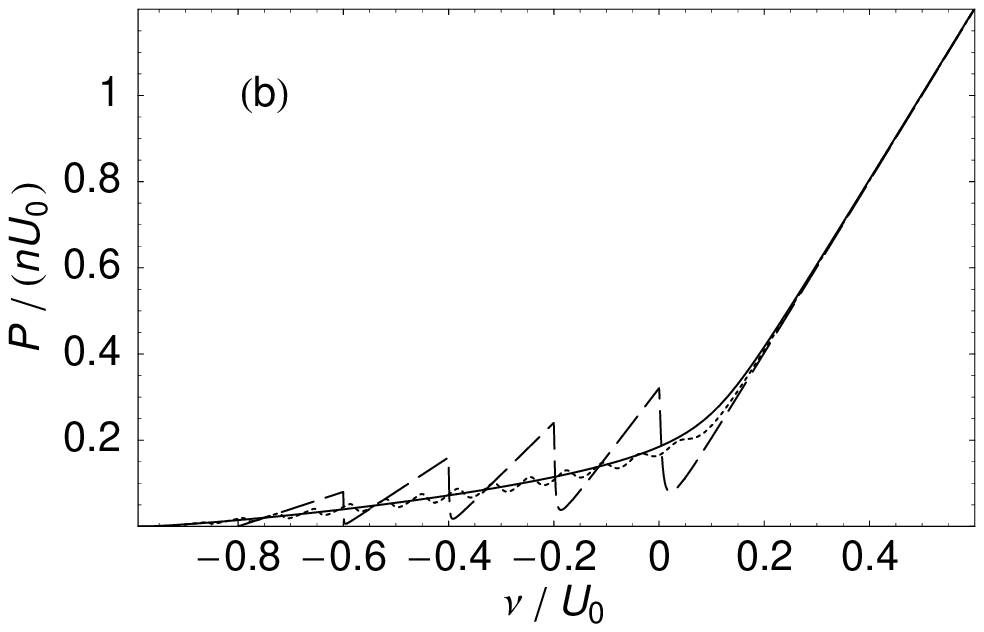}}
\caption{
\label{fig01:MC_T_and_p}
(a) Microcanonical temperature $T$ and (b) pressure $P$ as a function of
energy per particle $\nu=E/N$ for a (reduced) density
\mbox{$n=N/[L-(N-1)d_\mrm{hc}]=0.001/r_0$} and different number of particles $N=5$ 
(dashed line), $N=15$ (dotted), and $N=500$ (solid). Note that each of
the curves is $(N/2-2)$ times differentiable. 
}
\end{figure}

As shown in Fig. \ref{fig01:MC_T_and_p}, the {\em microcanonical}
caloric law~\eqref{e:caloric}  as well as the pressure $P(E)$ exhibit $N$
non-analytic points at the energies $E_k=-k U_0$, $k=0,\ldots,N-1$,
separating $N$ energetically different dissociation states (all
bindings intact, one binding broken, etc.). The formal
order~\cite{Eh12,Eh33,DuHi05} of these non-analyticities equals
\mbox{$N/2$}, i.e., the entropy has continuous derivatives up to order $(N/2-1)$, but the $(N/2)$th derivative becomes discontinuous  (a similar result was obtained recently by Kastner and
Schnetz for the mean-field spherical spin  
model~\cite{KaSc05}; see also Gross~\cite{Gr2004} for a general discussion of differentiability properties of the microcanonical partition function). Consequently,  the \lq microscopic (dissociation)
phases\rq\space as well as the singularities appear to be smoothened out in the
thermodynamic limit $N\to\infty$. Nevertheless, for finite
systems -- and in particular at small densities -- the non-analytic
behavior is accompanied by 
strong variations/oscillations of observable quantities as temperature
and pressure, when continuously varying $E$. Both qualitatively and quantitatively, this behavior is analogous to
what is usually denoted as a \lq phase transition\rq. However, since these
microscopic non-analyticities do not survive in the thermodynamic
limit (at least for our $1d$-model), they strictly speaking are not
covered by the conventional definition of \emph{singular macroscopic} PTs. We shall, therefore, speak of \emph{singular (or non-analytic) microscopic}
PTs in the MCE~\footnote{Due to the fact that the formal Ehrenfest order of the microscopic PTs increases with particle number, it is not particularly helpful to classify microscopic PTs according to this order.}. 
\par
Let us also briefly discuss the parameter dependence of the microcanonical TDFs shown in Fig.~\ref{fig01:MC_T_and_p}. As evident from Eq.~\eqref{e:result_omega-a}, the positions $E_k$ of the singular macroscopic PTs are just proportional to $U_0$.  The amplitude of the associated oscillations in $T$ and $P$ does also depend on the particle number $N$ and box size $L$: The strength of the oscillations increases for larger values $U_0$ and $L$, but becomes smaller for larger particle numbers $N$. The number and formal order of the PTs, however, only depend on the particle number $N$ and are independent of $U_0$ and $L$ (as long as $U_0>0$ and $L>L_\mrm{min}$). Thus, qualitatively, the results are independent of the particular choice of the model parameters $U_0$ and $L$. Moreover, analogous features can be found in the microcanonical caloric curves of $1d$ Lennard-Jones chains~\cite{DuHi05}.
\par
It should be mentioned that the exact phase volume~\eqref{e:result_omega-a} of our model system
resembles in structure the phase volume obtained by the Harmonic Superposition
Method~(HSM) applied to Lennard-Jones clusters (see, e.g., Doye \cite{DoyePhD96} or Wales and Doye~\cite{WaDo95} and
references therein). The HSM approximates the phase volume $\Omega(E)$ by a sum
of ellipsoidal regions around all local minima of the potential $U$ lower than
the total energy $E$. This method has been successfully applied to
describe \emph{melting phenomena}, as e.g. the low-temperature
properties of $3d$ Lennard-Jones clusters and their transition from a
solid-like state, where the cluster only vibrates around the ground
state configuration, to a liquid-like state, where also other locally
stable configurations energetically accessible.
The standard HSM approximation, however, does not properly account for the contribution to the phase volume
stemming from (partly) dissociated (or gas) states of the cluster, and
therefore, is not suitable for describing \emph{evaporation phenomena}. 
In particular, the HSM does not yield any singular microscopic PTs for $1d$ Lennard-Jones chains 
(where only one locally stable configuration, i.e. the ground state, exists) therewith contradicting exact analytical and numerical results~\cite{DuHi05}. By contrast, the model system discussed here -- if
considered as an approximation to Lennard-Jones chains -- does
reproduce these microscopic PTs related to evaporation (but, of course, our model cannot be applied to melting processes because it is one-dimensional). 
\par
It is worthwhile to discuss the microscopic PTs and the origin of the associated temperature oscillations, as observed in our model, from a more general point of view.
Mathematically, microscopic PTs of the above
type arise whenever the phase volume $\Go$ grows non-smoothly in the
vicinity of some critical energy value $E=E_k$. This can best be 
illustrated by considering the energetically admissible subset of the
\emph{configuration space}  
\be
\mcal A(E)=\left\{q\in \mbb R^N \;|\;\Theta(E-U(q))=1\right\}. 
\ee 
The set $\mcal A(E)$ consists of all positions space points
$q=(q_1,\ldots,q_N)$ that can be
occupied by the system at the given energy value $E$. Clearly, the
boundary of $\mcal{A}$, denoted by $\p \mcal{A}$, determines the
effective range of the integral in Eq.~\eqref{e:omega-1}. Hence,
whenever $\mcal A$ or $\p \mcal{A}$, respectively, change their
shape in an irregular (non-analytic) manner, a non-analyticity in the phase volume
$\Go$ may arise (and, hence, in the TDFs). For example, such a irregular
change in the shape of $\mcal{A}$ occurs when the energy for the next
dissociation step is crossed, since then some parts of the
boundary $\p \mcal{A}$ suddenly become determined by the box potential.  
\par
It remains to be discussed how the temperature oscillations -- i.e., the regions with negative heat capacity (also known as \lq S-bends\rq\space or van der Waals-type loops
\footnote{
Wales and Berry \cite{WaBe94} and Wales and Doye \cite{WaDo95} discuss necessary and sufficient criteria for the existence of S-bends in the microcanonical caloric curve based on the Boltzmann entropy, given by $S_B=\kB \ln (\p \Go/\p E)$ with respect to our notation. However, in general, Eq.~\eqref{e:average} does not hold for the temperature \mbox{$T_B=(\p S_B/\p E)^{-1}$} derived from the Boltzmann entropy, which is why we prefer to use the Hertz entropy \mbox{$S=\kB\ln \Go$}. The difference between Hertz entropy $S$ and Boltzmann entropy $S_B$ may become negligible in the thermodynamic limit, i.e., for sufficiently large systems, but is relevant for small systems.
})
 -- arise:
In the vicinity of the dissociation energy $E_k$, the set $\mcal A$ and, thus, also $\Omega$ and $S$ grow very rapidly, thereby giving rise to a drop-off in
temperature. Geometrically, this can be viewed as a sudden increase of
the `effective dimensionality' of $\mcal A$. Here, `effective
dimensionality' refers to the number of orthogonal configuration space
directions in which $\mcal A$ has an extent comparable to the system size
$L$. Hence, typically, the temperature oscillations  appear more pronounced for larger values of $L$.
From the physical point of view, the temperature decrease after the
$k$th dissociation step just means that for energy values slightly larger than
$E_k$ the dissociated fragments have very little kinetic energy (since
most of the energy has already been used to break the binding). The
larger the system the less likely it is that the fragments temporarily
recombine in a state of high kinetic energy; i.e., from a probabilistic
standpoint, the average \eqref{e:average} is then dominated by phase
space regions of low kinetic energy. With regard to practical applications, this means that \emph{one could cool such a small isolated system of bound particles by injecting energy until the fragmentation process sets in (cooling by evaporation/dissociation)}.
\par
The above described features of microscopic PTs are \emph{generic} and shared by all physical systems that exhibit dissociation and evaporation (e.g. similar
microscopic PTs and temperature oscillations are also found for small
$1d$-Lennard-Jones molecules \cite{DuHi05}). In particular, microscopic PTs should become more
pronounced in two or three dimensions (since then effective dimension
of $\mcal{A}$ grows even more rapidly at the dissociation levels) and
also be observable in quantum systems. By virtue of dissociation experiments with small particle numbers and very low
densities (similar to those of Schmidt et al. \cite{Sc01}, but without heat
bath), one should, in principle, be able to detect the oscillating
behavior e.g. in temperature and pressure curves. However, to actually
observe such oscillations one has to realize the requirements of the MCE,
i.e. a thermally isolated system with regulated energy
injection. Furthermore, due to the microscopic origin of the
oscillations and the requirement of a relatively low particle density,
a high sensitivity of the velocity and force measurements and long
measuring time spans will be necessary.

\par

\section{Canonical ensemble}
Employing the canonical ensemble (CE) is appropriate,
if the system under consideration is in thermal contact with a much larger
system
(heat bath), as e.g. realized in dusty cluster experiments
\cite{VaVlPeFo02}.
The relevant thermodynamic potential is the free energy~\cite{LaLi5}
\be
F(\beta,L,N)\equiv -\gb^{-1} \ln \mcal{Z}_\mrm{C}(\beta,L,N),
\ee 
where $\mcal{Z}_\mrm{C}$ is the canonical partition function. The external
control variables are now the inverse temperature
$\gb\equiv(\kBT)^{-1}$ of the heat bath, the volume $L$ and the particle
number $N$. For the above model, $\mcal{Z}_\mrm{C}$ can be exactly
calculated, analogous to Eq. \eqref{e:result_omega}, as 
\be\label{e:canon_result}
\mcal{Z}_\mrm{C}=\f{1}{N!}\left(\frac{2\pi m}{\gb h^2}\right)^{N/2}
\sum_{k=0}^{N-1}
\omega_k\;e^{\beta\,k U_0}
\ee
with $\go_k(N,L)$ given by Eq. \eqref{e:result_omega-b}. 
Mean energy and pressure of the CE are defined by  
$\bar{E}\equiv -{\p(\ln \mcal{Z}_\mrm{C})}/{\p\gb}$ and $\bar{P}\equiv
-{\p F}/{\p L}$,
yielding e.g. the {\em canonical} caloric law 
\be
\bar{E}=\f{N}{2\gb}-
\f{\sum_{k=0}^{N-1}\omega_k\;e^{\beta\,k U_0}\; kU_0}
{\sum_{k=0}^{N-1}\omega_k\;e^{\beta\,k U_0}}.
\ee
\par
\begin{figure}[tb]
\centerline{\includegraphics[width=1\linewidth]
{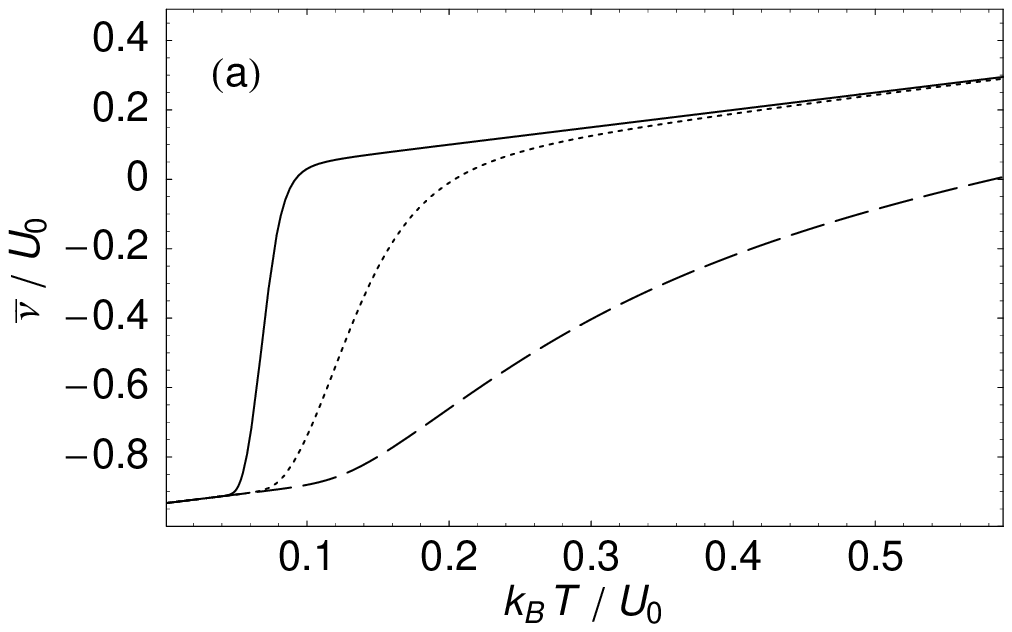}}
\centerline{\includegraphics[width=1\linewidth]
{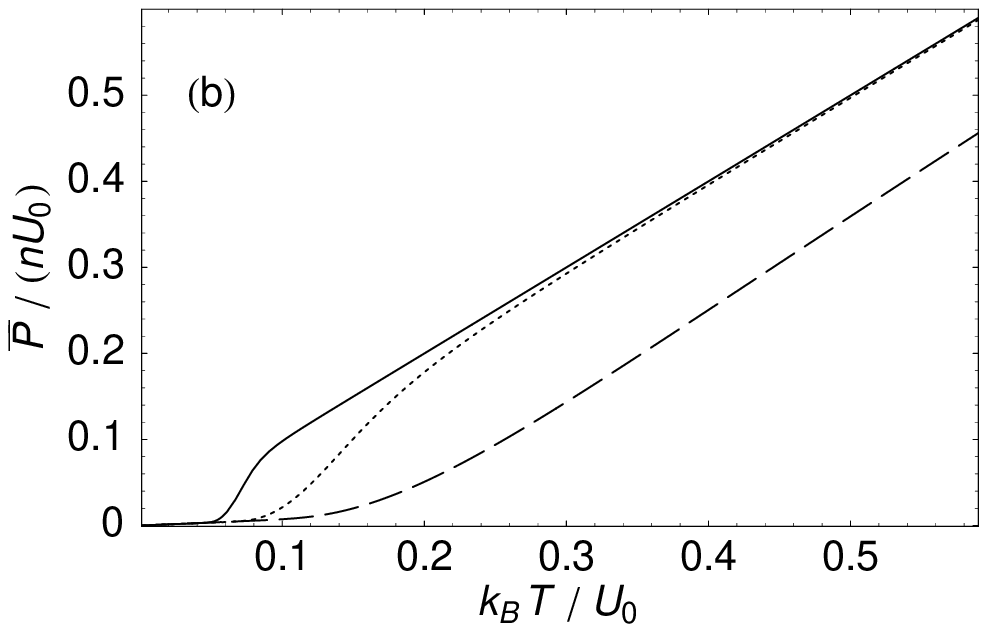}}
\centerline{\includegraphics[width=1\linewidth]
{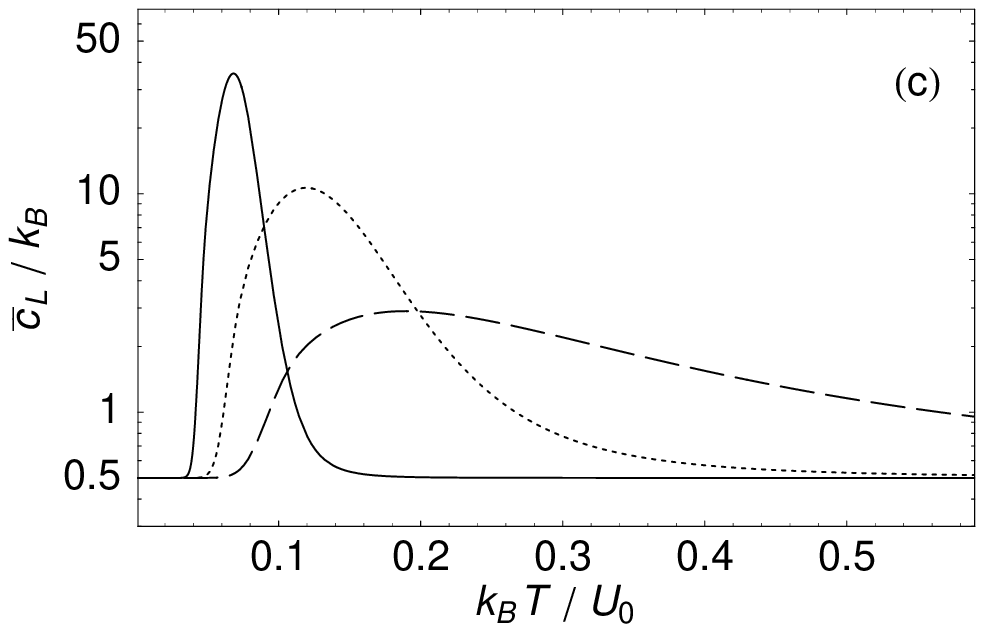}}
\caption{
\label{fig02:C_nubar_p_and_cL}
(a) Canonical mean energy per particle $\bar{\nu}=\bar{E}/N$, (b) pressure
$\bar{P}$ and (c) specific heat capacity $\bar{c}_L=\p \bar{E}/\p T$
(logarithmic scale)
as a function of temperature $T$ for $N=15$ particles and different values of
the reduced density
$n=10^{-1}/r_0$ (dashed line), $n=10^{-3}/r_0$ (dotted), and
$n=10^{-6}/r_0$ (full).    
} 
\end{figure}

Figures \ref{fig02:C_nubar_p_and_cL} (a) and (b) show $\bar{E}(T)$ and
$\bar{P}(T)$ for different values of the reduced particle density $\mbox{$n=N/[L-(N-1)d_\mrm{hc}]$}$. In contrast to the 
microcanonical pressure [Fig. \ref{fig01:MC_T_and_p} (b)], the canonical
pressure is a monotonous function of $T$ or $\bar{E}$, respectively. 
In the thermodynamic limit, microcanonical and canonical caloric 
curves become nearly indistinguishable. The canonical heat capacity
$\bar{c}_L=\p \bar{E}/\p T$
exhibits a strong (non-singular) peak in the temperature region, where 
dissociation occurs [Fig. \ref{fig02:C_nubar_p_and_cL} 
(c)]. If observed in an
experimentally measured curve, such behavior would possibly be interpreted as a
PT. For decreasing density $n$, the position of the maximum of $\bar{c}_L$
moves closer to $T=0$, while its height increases rapidly.
Furthermore, our results indicate that for $N\ge 15$ the TDFs $\bar{\nu}$ and
$\bar{c}_L$ become virtually independent of $N$. The (non-singular)
peak in the heat capacity persists in the thermodynamic limit
(analogous to the $1d$ Ising model~\cite{Sc00}). 
\par
To obtain a more detailed characterization of the dissociation process  
in the CE, we next study the
distribution of complex zeros (DOZ) of $\mcal{Z}_\mrm{C}$.
As evident from
Eq. \eqref{e:canon_result}, the only relevant configurational part of
$\mcal{Z}_\mrm{C}(\gb)$ is a polynomial of
degree $(N-1)$ in $z=e^{\gb U_0}$, and, therefore, has $(N-1)$ complex Fisher zeros \cite{Fi71} per branch of the logarithm. This quasi-polynomial structure is a consequence of the fact that, for our specific model, the configuration space $[0,L]^N$ can be partitioned into regions of equal total binding energy $E_k=-k U_0, k=0, \ldots,N-1$ (see Appendix). Since all zeros can be obtained by adding integer multiples of $2\pi i$ to the zeros $\gb_k$ of the main branch (for which $\Im\left(\gb_k\right)=\pi$), it suffices to discuss to the main branch only while bearing in mind that each zero $\gb_k$ in the main branch is associated with an infinite set of zeros $\{\gb_k+2\pi i\,s| s\in\Z\}$. Ordering the zeros according to their real parts, $\Re(\gb_0)\le \ldots \le\Re(\gb_{N-1})$, we find that the region of the
$\bar{c}_L$-peak is well-described by the temperature interval
$[\Re(\gb_{N-1})^{-1},\Re(\gb_0)^{-1}]$.
\par
\begin{figure}[tb]
\centerline{\includegraphics[width=1\linewidth]
{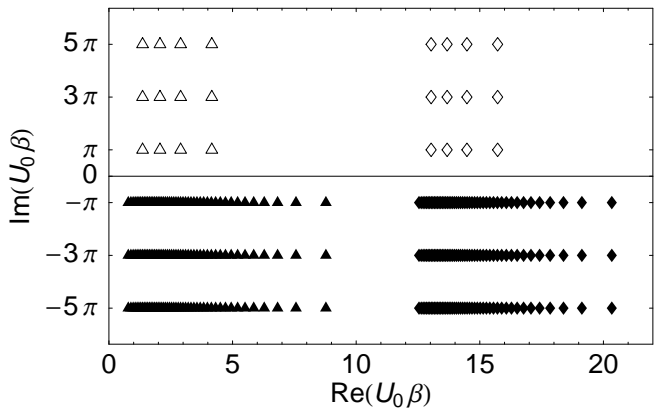}}
\caption{
\label{fig03:C_zeros}
DOZ of the canonical partition function
$\mcal{Z}_\mrm{C}(\beta)$ for $N=5$ (open symbols) and $N=15$ (filled symbols)
and $n=10^{-1}/r_0$ (triangles), and $n=10^{-6}/r_0$ (diamonds).
For better visibility, for $N=5$ (15) only the first few branches of
zeros with positive (negative) imaginary part are shown (further
branches can be obtained by shifting with multiples of $2\pi i$ in the
vertical direction; generally, the Fisher zeros are symmetric with
respect to the real axis \cite{BoMuHa00}).
} 
\end{figure}

The asymptotic behavior of the DOZ for $N\to\infty$ may be used to characterize the parameters (critical temperature, order, etc.) of singular
macroscopic PTs \cite{YaLe52,LeYa52,Fi71,GrRo67,GrRo69,GrLe69}. In our model, the Fisher zeros are located at least a distance $\pi$ away from the the real $\gb$-axis regardless of particle number $N$ (see Fig.~\ref{fig03:C_zeros}). Hence, the zeros cannot converge to a point on the real $\gb$-axis. The peculiar position of the zeros thus ensures agreement with the (generalized) van Hove theorem~\cite{Ho50,CuSa04} which excludes the existence of a singular macroscopic PT in our model.
\par
The DOZ has also been employed to study their finite size analogs of macroscopic PTs \cite{BoMuHa00,MuStBo01,JaKe01,AlFeHa02}. Applying these methods to our model, one may interpret the smooth phase transition observed in the CE as a superposition of $(N-1)$ smooth microscopic `first-order' phase transitions indicated by the $(N-1)$ sets of zeros $\{\gb_k+2\pi i\,s| s\in\Z\}$. 
One can then use $\{\Re(\gb_0), \ldots, \Re(\gb_{N-1})\}$ to distinguish (define) different dissociation states in the CE.
Thus, our results suggest that the DOZ encodes detailed information
about the observed smooth phase transition and the energetically
different degrees of dissociation even if 
there is no singular PT in the thermodynamic limit. In particular, since
microcanonical phase volume and canonical partition function can be mapped onto
each other via the Laplace transformation \cite{PeHaTi85,MuelkenPhD01}, one may
speculate that there exists a direct mathematical link between the
singular microscopic PTs in
the MCE and the DOZ in the~CE.

\section{Summary} 
We have studied a simple $1d$-evaporation model with nearest neighbor
interaction potentials characterized by a hard-core repulsive part and
piecewise constant short-range attraction. By analyzing the
exact TDFs of this model, it was shown that, in the case of a thermally isolated
system (MCE), the
microcanonical caloric and pressure laws exhibit singularities, separating 
different dissociation states. The formal order of these non-analyticities
increases as the particle number increases. Hence, the microscopic PTs
vanish in the thermodynamic limit and are intrinsically different from the
topologically induced macroscopic PTs discussed in Refs.
\cite{CaCoPe02,FrPe04,AnEtAl05,Kastner04,TeSt04}.
\par
For sufficiently low particle numbers and densities, the microscopic PTs are accompanied by strong
oscillations of the temperature (mean kinetic energy) and pressure.
These oscillations arise from a rapid change of the phase volume near the dissociation thresholds. They are a generic feature of particle systems with Lennard-Jones like interaction potentials~\cite{DuHi05,Gr2004}. They are not restricted to one space dimension, but expected to be even stronger in two and three space dimensions, and should also be observable in quantum systems. In a suitably designed dissociation experiment, one should therefore, in principle, be able to detect the oscillating behavior e.g. in temperature and pressure curves. In particular, such temperature oscillations may provide the possibility to cool a small isolated system by means of regulated energy injection (cooling by evaporation).
\par
If the model system is coupled to a heat bath (CE), a smooth PT is observed, but
no singularities are found \cite{Ho50,CuSa04}. Nevertheless, the DOZ seemingly
permits to quantify the temperature range of the PT and to define different
dissociation states.
\par
We thus conclude this paper with two important questions, which
need to be answered in the future:
Is it possible to design an experiment that allows to observe singular
microscopic phase transitions in finite size systems (or, at least, oscillatory
behavior of thermodynamic observables such as pressure), as predicted by the
microcanonical statistical theory? Can one find a direct mathematical link
between singularities in microcanonical partition function of finite systems and
the DOZ of the corresponding canonical partition function -- and, thus, between
microscopic and macroscopic phase transitions in arbitrary space dimensions? 

\begin{acknowledgments}
The authors would like to thank M.~Kastner, L.~Velazquez-Abad and an anonymous referee for several helpful remarks. 
\end{acknowledgments}

\bibliographystyle{apsrev}
\bibliography{TD}

\appendix
\section*{Appendix} To calculate integral \eqref{e:omega-1}, we
first eliminate the hard-core part of the interaction potential by virtue of the
transformation $L \mapsto \gl=L-(N-1)d_\mrm{hc},\;
q_i \mapsto x_i = q_i- n_i d_\mrm{hc}\in[0,\gl],$
here $n_i$ equals the number of particles~$j$ with \mbox{$q_j<q_i$}.
With these definitions the potential energy can be rewritten as:
\bse
\begin{equation}\notag
U'(\mbf{x};\gl,N)= \label{e:new_hamiltonian}
\f{1}{2}{\sum_{i,j}}^\prime
U'_\mrm{pair}(|x_i-x_j|)+
U_\mrm{box}(\mbf{x};\gl,N)\;,
\end{equation}
where the sum $\sum^\prime$ goes over nearest neighbors only, and
\begin{align}\notag
U'_\mrm{pair}(r) =& \begin{cases}
-U_0,& 0<r<r_0,\\
0, & r\geq r_0\;.
\end{cases}
\end{align}
\ese
For $k=0,\ldots,N-1$ the family of sets \mbox{$\mcal{G}_k=
\left\{\mbf x\in [0,\gl]^N\;|\;U'(\mbf x)=-k U_0\right\}$} constitutes a
partition (disjoint cover) of the configuration space
permitted by the transformed box volume $[0,\gl]$. Thus we can rewrite
Eq. \eqref{e:omega-1}  in the form 
\be
\Omega\notag
=C(N) \label{e:omega_integral-2}
\sum_{k=0}^{N-1}
(E+kU_0)^{N/2}\;\Theta(E+k U_0)\;\vol(\mcal{G}_k).
\ee
In order to calculate $\vol(\mcal{G}_k)$ we first note that
\be
\go_k\equiv \vol(\mcal{G}_k)&=&\notag
N!\cdot\vol(\mcal{G}_k^+),
\ee
where (writing \mbox{$\left\{\mbf x\in
[0,\gl]^N\;|\;\ldots\;\right\}=\left\{\ldots \right\}$ from
now on}) 
\be
\mcal{G}_k^+=\notag
\left\{x_1<x_2<\ldots<x_N\;\wedge\;U(\mbf q)=-k U_0\right\}.
\ee
Next we note that
\be\notag
\vol(\mcal{G}_k^+)=\binom{N-1}{k}\cdot\vol(\mcal{G}_k^{++}),
\ee
where
\be
\mcal{G}_k^{++}&=&\notag
\left\{x_1<x_2<\ldots<x_N\right\}\cap\\
&&\notag
\;\bigl\{(x_2-x_1<r_0)\wedge\ldots\wedge (x_{k+1}-x_k<r_0)\wedge\\
&&\notag
\;\;\;(x_{k+2}-x_{k+1}\ge r_0)\wedge\ldots\wedge(x_{N}-x_{N-1}\ge r_0)\bigr\},\\
\vol(\mcal{G}_k^{++})&=&\notag
\int_{(N-1-k)r_0}^\gl \diff x_N\int_{(N-2-k)r_0}^{x_N-r_0} \diff x_{N-1}
\times\cdots\\
&&\label{e:g++}\notag
\int_{r_0}^{x_{k+3}-r_0} \diff x_{k+2}\int_{0}^{x_{k+2}-r_0} \diff x_{k+1}
\times\\
&&\notag
\int_{[x_{k+1}-r_0]_+}^{x_{k+1}} \diff x_{k}\cdots
\int_{[x_3-r_0]_+}^{x_{3}} \diff x_{2}
\int_{[x_2-r_0]_+}^{x_{2}} \diff x_{1}.
\ee
The positive part $[x]_+\equiv\max\{0,x\}$ satisfies
\bse
\begin{align}
\int\diff x\;[x]_+^n &=\f{[x]_+^{n+1}}{n+1},&n\in \mbb{N};\\
\left[[x-c\right]_+ -c]_+&=[x-2c]_+,        &c>0.
\end{align}
\ese
By virtue of these identities, we can write Eq. \eqref{e:g++} as
\be
\vol(\mcal{G}_k^{++})&=&\notag
\int_{(N-1-k)r_0}^\gl \diff x_N\int_{(N-2-k)r_0}^{x_N-r_0} \diff x_{N-1}
\times\cdots\\
&&\notag
\int_{r_0}^{x_{k+3}-r_0} 
\diff x_{k+2}\int_{0}^{x_{k+2}-r_0} \diff x_{k+1}\; K_k,
\ee
where
\be
K_{k}
&=&\notag
\f{1}{k!}\sum_{i=0}^k\;
(-1)^{i}\;\binom{k}{i}\;[x_{k+1}-i r_0]_+^k.
\ee
Performing the remaining integrations and reversing the
transformation $L\mapsto \gl$, one obtains the final result
\eqref{e:result_omega-b}.

\end{document}